# Multilateralism in the Global Governance of Artificial Intelligence

Pre-print version

**Michal Natorski**
**Maastricht University/UNU-MERIT**

**https://orcid.org/0000-0002-2736-1709**

**micha.natorski@maastrichtuniversity.nl**

Michal Natorski is an Assistant Professor at Maastricht University and a Research Associate at the United Nations University-MERIT in the Netherlands.

## Abstract

This chapter inquires how international multilateralism addresses the emergence of the general-purpose technology of Artificial Intelligence. In more detail, it analyses two key features of AI multilateralism: its generalized principles and the coordination of state relations in the realm of AI. Firstly, it distinguishes the generalized principles of AI multilateralism of epochal change, determinism, and dialectical understanding. In the second place, the adaptation of multilateralism to AI led to the integration of AI issues into the agendas of existing cooperation frameworks and the creation of new ad hoc frameworks focusing exclusively on AI issues. In both cases, AI multilateralism develops in the shadow of the state hierarchy in relations with other AI stakeholders. While AI multilateralism is multi-stakeholder, and the hierarchy between state and non-state actors may seem blurred, states preserve the competence as decisive decision-makers in agenda-setting, negotiation, and implementation of soft law international commitments.

## 1. Introduction

Artificial Intelligence (AI) is considered a general-purpose technology. General-purpose technologies have transformative, long-standing effects on economies and societies due to their singular, recognizable, and generic technological features, which are widely used, have multiple applications, and generate numerous spillover effects (Lipsey et al., 2005). Such general-purpose technologies as money, weapons, the steam engine, electricity, and computers exemplify the multiplicity of scope, breadth, and duration of uses and effects of technological advancements. Technological changes altered actors' preferences and status, leading to social, political, and economic disruptions. They are underlying factors in the redefinition of normative and institutional forms of international relations (Drezner, 2019). Therefore, it can be expected that AI as a general-purpose technology will affect directly and indirectly all areas of human activity and will also influence the traditional international institutions of diplomacy (Bano et al., 2025), sovereignty (Scasa, 2023), and war (Johnson, 2024).

Technological advancements in AI have sparked debates about the architecture of global governance and the mechanisms for regulating AI. Given its prominent role on the international agenda, it has become an issue addressed by many multilateral collaboration frameworks worldwide. One of the most comprehensive efforts to assess the emerging architecture of global governance of AI is the report "*Governing AI for Humanity",* prepared by the High-level Advisory Body on Artificial Intelligence of the United Nations Secretary-General (United Nations AI Advisory Body, 2024). This report acknowledged the multitude of existing initiatives to govern AI but also emphasized that despite "hundreds of guides, frameworks, and principles adopted by governments, companies, consortiums, and regional and international organizations" (Ibidem: 8), there is no comprehensive and representative global framework for coordination and implementation of AI global policy (Roberts et al. 2024; Geith et al. 2025). Only a few states actively and continuously participate in the most significant international collaboration on AI governance, resulting in a deficit in AI global governance in many parts of the world.

To explore the paradox of the simultaneous multiplicity of frameworks and the deficit of participation in AI global governance, this chapter asks: How did international multilateralism address the emergence of the general-purpose technology of Artificial Intelligence? At the beginning of the recent AI hype, there was an expectation that "multilateral organizations could play a pivotal role in AI governance by providing a joint forum for the formulation, coordination and dissemination of the cooperative norms between actors, enabling participating to signal sincere commitment for beneficial and shared AI development" (Dafoe, 2019: 122). While at that moment there was no organization meeting such expectations, it is an empirical but also analytical question to understand how multilateralism reacted and adapted to AI afterwards.

A classic definition of multilateralism is the coordination of relations among three or more states in accordance with certain "generalized" principles "which specify appropriate conduct for a class of actions" (Ruggie, 1992: 571). Consequently, this chapter analyses two key features of multilateralism: the generalized principles and the coordination of state relations in the realm of AI. The subsequent step is to analyze how AI-related generalized principles influenced the role of states and the coordination of their relations. How did the existing multilateral framework adapt to the AI topic? Conversely, has a new multilateral framework for AI governance emerged? How do these processes affect the role of states in AI multilateralism? The answers to these questions will indicate whether AI as a general-purpose technology has transformed multilateralism as a form of coordination among states.

To address these issues, the argument developed in this chapter points out that AI, as a general-purpose technology, is framed by a distinctive discourse legitimizing a particular conduct of states' actions. The principles of epochal change, determinism, and dialectical understandings concerning AI governance subsequently led to the adaptation of multilateralism, as AI issues were integrated into the agendas of existing cooperation frameworks and new ad hoc frameworks were established, focusing exclusively on AI issues. In both cases, AI multilateralism develops in the shadow of the state's hierarchical authority in relations with other AI stakeholders (Scharpf, 1997). While AI multilateralism is multi-stakeholder, and the hierarchy between state and non-state actors may seem blurred, states preserve the competence as decisive decision-makers in agenda-setting, negotiation, and implementation of soft law international commitments.

This chapter draws from publicly available information, direct observation, and participation in UN system and other frameworks debates (AI for Good Summits in Geneva and Responsible Artificial Intelligence in the Military Domain (REAIM) summits) until mid-2025. It is organized into five sections. Following this introduction, the subsequent section provides a brief overview of the literature on AI global governance and introduces the concept of multilateralism. Subsequently, the chapter outlines the core characteristics and organizing principles of discourse legitimizing AI multilateralism. The following three sections characterize the complexity of the emerging AI global governance frameworks while emphasizing the leading role of states as hierarchically superior actors in governing this general-purpose technology. The final section summarizes the findings and outlines some follow-up directions of the research agenda regarding AI global governance and multilateralism.

## 2. Multilateral Perspective and AI

This section introduces the topic of AI multilateralism in two analytical steps. The first part outlines the prominent strands of academic work on AI from a conceptual perspective of global governance. The second part highlights key elements of the concept of multilateralism. This contrast helps distil the analytical question of the state's role in AI global governance from the conceptual perspective of multilateralism.

### *2.1. The academic cartography of AI global governance*

AI has become a topic of numerous international initiatives within existing international organizations and the institutionalization of new frameworks. In academic work, these international initiatives are scrutinized under the broad umbrella of (global) governance; however, it is not always clearly conceptualized from the outset.

Following the approach of global governance, different categories were proposed to map the rapidly changing landscape of AI global governance. For example, Butcher and Beridze (2019) provided an overview of early initiatives. They distinguished them by the varying levels of involvement of the private sector, public sector, non-governmental organizations, research institutions, multi-stakeholder organizations, and the United Nations (UN). Schmitt (2022) categorized the emerging AI governance initiatives around the distinction between the initiatives within the existing governance frameworks and newly established AI-specific frameworks and the leadership emanating from states and other non-state actors, such as the International Organization's bureaucracies or private actors. Tallberg *et al.* (2023) mapped AI initiatives around not mutually exclusive categories of regulation scope (horizontal vs. vertical), authority (centralized vs. decentralized), commitments (soft law vs. hard law), origin (private vs. public), and subject matter (security-military vs. non-military). Finally, Vaele *et al.* (2023) distinguished the following six modalities of AI's global governance commitments: ethical codes and councils (public and private), industry self-governance, contracts and licensing, standards, international agreements, and extraterritorial domestic regulations.

Besides these early attempts at categorizing empirical analyses, the emerging architecture of AI global governance has been studied predominantly from a normative perspective, focusing on desirable institutional frameworks. Scholars proposed to anchor AI debates and regulations mirroring different international frameworks, such as the International Panel on Climate Change, the International Atomic Energy Agency, or CERN, to establish, for example, the Intergovernmental Panel on AI, the International AI Organization, the UN AI Control Agency, or the UN AI Research Organization. Maas and Villalobos (2023) mapped 47 proposals for AI global governance institutions across seven dimensions of possible organizations, focusing on scientific consensus-building, political consensus-building, and norm-setting, coordination of policy and regulation, enforcement of standards or restrictions, stabilisation and emergency responses, international joint research, and distribution of benefits and access. On the other hand, Ho *et al.* (2023), given the irruption of frontier AI, proposed to create four international bodies to establish scientific positions on advanced AI, set governance norms and standards, promote access to advanced AI, and mitigate AI risks by promoting its safety. There are also different proposals to apply or adapt existing international institutions and norms to the challenges of AI (Cihon et al., 2020). Consequently, many analyses of the EU's efforts to regulate AI focus on the EU AI Act, the first international legally binding regulation of AI (Almada, 2025). There are also other academic analyses of AI debates within the existing framework, such as those by UNESCO (Natorski, 2024), the ITU (Leander, 2021), and the United Nations (Teleanu, 2024).

Besides these broad trends, research on AI governance relevant to IR has focused on a few topical issues. Firstly, building on the variety of actors and commitments surrounding the emerging regulatory landscape of AI, numerous studies have mapped the nature of ethical principles and guidelines across both the private and public sectors. For example, Jobin et al. (2019) observed substantive differences in the interpretation, relevance, applicability, and implementation of the five core principles of transparency, justice and fairness, non-maleficence, responsibility, and privacy (see also Hagendorff, 2020). Another mapping of AI ethical principles corroborated most of the above findings on the relevance of ethical principles, but added the principles of reliability and safety to the priorities. Private industry and governmental institutions dominated this global mapping, while international organisations were less significant (Kluge Correa *et al.*, 2023). In the second place, another strand of research on AI in the international context emphasized the global debates on AI in the military domain. Firstly, scholars have highlighted the challenges in reaching agreements during debates on Lethal Autonomous Weapons Systems (LAWS) within the framework of the UN Convention on Certain Conventional Weapons (UN CCW) (Bode and Huelss, 2018; Bode, 2023; 2024). Similarly, the application of AI in defence and its consequences for global security focused on the significance of AI-propelled military power in such classic IR concepts as deterrence (Jensen *et al.,* 2020), the balance of power (Horowitz, 2018), confidence-building measures (Puscas, 2023, 2024), and arms race (Haner *et al.,* 2019).

The above examples of mapping AI international initiatives point out a multiplicity of existing and emerging relevant frameworks and areas for research under a broad conceptual umbrella of global governance. This loosely defined concept highlights regulatory processes involving multiple actors, blurring the distinction between public authority and private initiative. Typically, global governance emphasizes the institutionalized components of coordination among non-hierarchically organized actors, resulting in a collective commitment. It suggests the continuous transformation of the roles of state and non-state actors in the global processes (Domínguez and Velázquez Flores, 2018, Chapter Talberg, 2025). Drawing from this concept, it can be inferred that in a "multi-stakeholder" AI global governance process, the hierarchical role of the state is equivalent to that of non-state actors. To scrutinize this assumption, this chapter adopts the perspective of multilateralism as a specific form of AI global governance by emphasizing the interactions among states.

### *2.2. Multilateralism as an international institution*

Multilateralism is the coordination of relations among three or more states in accordance with certain "generalised" principles, "which specify appropriate conduct for a class of actions" (Ruggie, 1992: 571). Multilateral relations are governed by agreed-upon rules and principles that constrain the choices of actors (Ikenberry, 2003: 534). Hence, multilateralism is a characteristic of international institutions based on ordering principles that shape expectations and constraints regarding the appropriate conduct of activities. While it is usually focused on the relations between states, it can also involve other actors in the frameworks regulated by states and international organizations (Cox, 1992). The involvement of numerous non-state private actors, including multinational corporations and civil society non-governmental organizations, has led to the proliferation of various public-private frameworks in multilateral diplomacy (Muldoon and Aviel, 2020).

Multilateralism also refers to a particular form of diplomatic practice. Multilateral negotiation is an instance of the broad institution of multilateralism. Multilateral negotiations involve multiple actors, interests, complex agendas, and differentiated international settings. In such contexts, negotiations involve coalition-building among states, nonstate actors, and international organizations (Hampson with Hart, 1999). The debates about multilateralism and its manifestations also emphasize its evolving character. Therefore, Pouliot (2011) emphasized also that multilateralism is a specific governance practice characterized by an inclusive, institutionalized, and principled form of political dialogue differentiated by the comprehensiveness, routinization, and non-discrimination from other forms of state relations such as unilateralism, bilateralism, minilateralism, coalitions of the willing, or concert-type organizations. The practices of multilateralism are sustained by legitimizing discourses, which also influence their forms and manifestations. As a diplomatic practice, multilateralism focuses on repetitive and patterned actions of state agents in the institutionalized context of global governance.

After the end of the Cold War, multilateralism has been attributed to a stabilizing role in the context of geopolitical shifts in Europe and the world (Ruggie, 1992). Multilateralism is also considered an outstanding international institution that characterizes the International Liberal Order (Ikenberry, 2015). However, it is widely observed that the Liberal Order seems to be in decline, hence its institutional form of multilateralism. As argued by Mansfield and Rudra (2021), the global digital interdependence has led to the erosion of embedded liberalism in the post-Second World War order, driven by an unprecedented increase in economic transactions between developed and developing countries. It leads to the erosion of support for multilateralism in the advanced industrial world, given the accumulation of distributional losses in these economies and an increasingly uncontrolled competition for capital between multinational companies and industrial outsourcing (Mansfield and Rudra, 2021). At the same time, it is also challenged by the rise of powerful information and communications technology (ICT) companies, which undermine the power of the most powerful states (Schaake, 2024) and shape the global order in an unprecedented way (Bremmer, 2021). Therefore, multilateralism may lose support among its previous promoters from developed economies as a potential solution to the surge of AI. The question is how multilateralism addresses the rise of a new general-purpose AI technology, which is expected to play a fundamental role in reshaping the global security, political, economic, and social landscape, given the redistribution of power, resources, and structures.

## 3. Organising principles of multilateralism for AI governance

Based on the above overview of the concept of multilateralism, the first task is to define the organising principle(s) that shape actors' actions related to AI. Such organized principles emerge from discursive strategies that normatively define AI technology in multilateral contexts. It is argued here that the collectively agreed-upon outcomes of multilateral processes, determine the multilateral organizing principles. To define them, I briefly characterize three common discursive strategies of so-called AI-essentialism (Schiølin, 2020; Kim, 2023) that, in practice, guide the embedding of AI technology in multilateral frameworks. They organize the multilateral practices around three principles: 1) epochalism - AI technologies open a new temporal era disrupting historical trajectories, 2) determinism - AI is inevitable, and 3) dialectics – AI implies dialectical understandings embracing contradictions. They shape the emerging features of AI governance and influence the nature of multilateralism concerning general-purpose AI technology. To illustrate them, I draw on public documents agreed upon in multilateral frameworks, such as the UN system, the OECD, the G7, the Council of Europe, REAIM, and NATO.

The first trait of the multilateral discourse on AI is its epochalism, which emphasizes the respect for social and technological advancements over essentialist reverence for traditional values (Geertz, 1973: 243-249). Epochalism emphasizes a radical change in human history due to technological advances, constituting a new era of development based on techno-solutionism, claiming that all societal problems can be solved by technology (Morozov, 2013). The emphasis on radical revolutionary technological change makes predicting their future consequences uncertain. The UN Global Digital Compact annexed to the resolution Pact for the Future (United Nations, General Assembly, 2024a) opens with urgency and exceptionalism, affirming that "digital technologies are dramatically transforming our world." It was recalled in the Africa Declaration on Artificial Intelligence, recognizing its "transformative potential" and "the exponential pace of development and adoption" (Global AI Summit on Africa, 2025). In the military realm, the REAIM Call for Action preamble states that AI, in general, "is influencing and changing our world fundamentally." As such, it "will drastically impact the future of military operations, just as it impacts the way we work and live" (REAIM, 2023). In such framings, AI is considered unique in human history; consequently, past solutions are no longer applicable. The traits of such a unique and exceptional era of AI justify the need to develop new special rules and frameworks to govern AI, or to renovate the existing ones to face the unique and unprecedented reality of AI.

This epochal framing is further reinforced by the understanding that the AI era is inevitable. It is argued that AI is already widely employed in various spheres of life. For example, the Pact of the Future states that "Digital and emerging technologies, including artificial intelligence, play a significant role as enablers of sustainable development and are dramatically changing our world. They offer huge potential for progress for the benefit of people and planet today and in the future" (United Nations, General Assembly, 2024a). The Recommendation adopted by OECD (2019) state that "AI has pervasive, far-reaching and global implications that are transforming societies, economic sectors and the world of work". Such transversal effects of AI "are likely to increasingly do so in the future"; therefore, they become inevitable. AI is framed as omnipresent in all spheres of life. Such technological determinism requires actions "to save the future and make it desirable" (Schiølin, 2020).

Despite the inevitability of the AI era, the technology is framed by uncertainties about the influence of AI, leading to omnipresent dialectics of optimism and pessimism, risks and opportunities, and innovation and regulation (Kim, 2023; Bradford, 2023; Tallberg et al., 2024). As a result, the key frame of the AI debates is constituted by juxtapositions of opportunities and risks. For example, UN resolution on AI for sustainable development stressed that the advance of AI "has the potential to bring new opportunities for socioeconomic development and accelerate the progress and achievement" of the SDGs, but "the improper or malicious design, development, deployment and use of" AI systems "could pose potential risks and challenges" (United Nations General Assembly, 2024b). Similarly, the AI Safety Summit declaration balanced the affirmation that AI "presents enormous global opportunities: it has the potential to transform and enhance human wellbeing, peace and prosperity" with the warning about "significant risks" and, in particular, "safety risks" of general purpose AI in areas such as cybersecurity, biotechnology and disinformation given a possible "serious, even catastrophic, harm" (AI Safety Summit, 2023). As a result, such a dialectical understanding, given the inevitable new AI era, implies an alternative between actions to explore opportunities and avoid risks or inaction, which will still lead to inevitable, likely negative consequences.

While the traits of epoch-making, determinism, and dialectical understandings are common across different governance frameworks, the considerations about their practical consequences are unsettled. These organizing principles of AI global governance suggest two effects of AI on multilateralism. Firstly, the epoch-making and deterministic inevitability suggests a shift in AI multilateralism, either through the adaptation of the existing multilateral framework or the emergence of a new multilateralism adapted to the realities of general-purpose AI technology. One way to adapt would be to involve all subjects affected by AI in the multilateral process, beyond state representatives. Therefore, given that most AI governance frameworks include a "multi-stakeholder" component to allow broad participation, representing different voices and positions beyond state control, such democratization would suggest that states would play an equal role alongside other stakeholders.

Second, the dialectical understanding of AI consequences also leads to the emergence of a contradictory approach to adapted multilateralism. On the one hand, AI can be seen as a catalyst for renewed multilateralism. On the other hand, it can be presented as an excuse to restrict multilateral cooperation. For example, during the negotiations on the UNECSO Recommendations on Ethical AI, participating countries adopted a wide range of positions, combining communitarian/sovereignist and cosmopolitan/multilateralist views on the ethical regulation of AI globally (Natorski, 2024). The distinctive feature of debates was the cleavage about the expected role of states in the AI governance framework. While many countries preferred to maintain the exclusive prerogatives of individual states within restricted international collaboration, others emphasized the need for universal, multilateral cooperation among states. Despite these differences, states continue to play their relevant roles by enabling or constraining the expansion of multilateral collaboration in this new area.

The following two sections analyze the adaptation of multilateralism to govern AI by contrasting the involvement of stakeholders with the role of states. In the next two sections, I examine how multilateralism adapts to govern AI technology by expanding existing international frameworks and organizations, or establishing new, flexible multilateral frameworks among mostly like-minded states. The following section analyzes the universalization of multilateral cooperation on AI within the framework of the UN system. In both cases, AI multi-stakeholder processes are developed in the shadow of the state hierarchical authority.

## 4. Membership in restricted multilateral frameworks

A prominent strand of AI governance has emerged within restricted multilateral frameworks, operating in the shadow of the state's hierarchical authority. While all processes involve different stakeholders, the dominant role of states is reflected in the rules governing core and expanded membership, as well as in their ultimate authority over the agenda, the negotiations of commitments, and the responsibility for implementation. In this context, AI multilateralism emerged due to two parallel processes of expansion of existing ones or the establishment of new frameworks (see Table 1).

**Table 1. Types of AI multilateralism**

| | **Existing framework** | **New framework** |
|---|---|---|
| **Open membership** | • G7 Hiroshima Process<br>• OECD Recommendations<br>• Council of Europe Convention<br>• UNESCO Recommandations<br>• UN CCC<br>• UN General Assembly resolutions | • GPAI<br>• REAIM<br>• Global AI Governance Initiative<br>• AI Summits<br>• Political Declaration on the Responsible Military Use of Artificial Intelligence and Autonomy |
| **Closed membership** | • LAC Council<br>• AU Strategy<br>• EU AI Act | • AI Partnership for Defence |
| | • NATO Strategy<br>• ASEAN Guidelines | |

Source: Own elaboration

The existing thematic and/or regional multilateral frameworks are expanding their *de facto* mandates to include AI governance in their activities. The restricted framework for AI governance depended on two pre-existing membership rules of international organizations (IOs) attributing the leading roles to the governments: 1) the advanced economic status in the global economic order (G-7, G-20, OECD), or 2) the geographical position of countries within world regions (EU, Council of Europe, African Union (AU)). The existing IOs and frameworks have approached AI governance as another strand of their activities.

The first type of restricted membership, based on economic status, was relatively open to voluntary universalization regarding AI commitments, allowing the participation of interested, like-minded non-member states. This differs from the closed mini-lateralism and the coalitions of the willing form of collaboration. An overview of these processes emphasizes the economic dimensions of digital transformation and AI. Therefore, the G-7 and OECD, as global Western organizations symbolising the Liberal International Order, played pioneering roles in discussing AI within the existing restricted multilateral frameworks.

For example, the OECD Recommendation has 47 adherents (38 members and 9 non-members in 2024), including non-members such as Brazil, Argentina, Ukraine, and the European Union. The 2024 G7 Hiroshima Process launched the Friends Group for states, as well as the Friends Group Partners' Community for other stakeholders. As a result, Friends Group includes 54 states and the European Union, representing G-7 members, Latin America (e.g. Argentina, Chile, Costa Rica), Asia (e.g. India, Laos, Vietnam, South Korea, Singapore), Africa (e.g. Kenya, Nigeria), Middle East (e.g. Israel, United Emirates Arabs) and Europea (many EU member states and also Norway and Turkey). The Friends Group Partners' Community comprises industry leaders (Amazon, Microsoft, Rakuten, OpenAI), international organizations (OECD, UNDP, World Bank), and civil society organizations (CAIDP, Safer AI). This open community is a voluntary framework of countries that supports the spirit of the Hiroshima AI Process in building global AI governance (Hiroshima AI Process, 2025). Similarly, the negotiation and adherence to this Council or Europe's Framework Convention on artificial intelligence, human rights, democracy and the rule of law visualized its global ambition going beyond the regional scope; it was drafted by 46 members of the Council of Europe, all observers, including the United States, Canada, Japan, Mexico, the Holy See, and the European Union, and six non-members states (Australia, Argentina, Costa Rica, Israel, Peru, and Uruguay). Therefore, this framework, based on the Council of Europe's standards on human rights, democracy, and the rule of law, aims to become a global standard open to the adherence of all states (Council of Europe, 2024).

However, many regional multilateral commitments are restricted to the participating governments, given that their membership rules emphasize territorial criteria. This is reflected in regional commitments, such as those of the AU, NATO, ASEAN, and the EU, which are primarily restricted to member states, despite their rules potentially impacting organizations from third countries. The African Union (AU) adopted in 2024 a Continental AI Strategy (African Union, 2024) that calls to action various actors, including the AU Commission, AU Member States, the private sector, and development partners. However, it remains clear that member states play a crucial role, which should, among other things, involve developing their national strategies and governance mechanisms for AI. Similarly, the Latin American and Caribbean countries organized two Ministerial summits (Santiago de Chile in 2023 and Montevideo in 2024) on the Ethics of Artificial Intelligence in Latin America and the Caribbean. The first summit decided to establish an intergovernmental Working Group with "the aim of creating an intergovernmental Council on Artificial Intelligence for Latin America and the Caribbean" (Declaración de Santiago, 2023). The second summit adopted a roadmap to "strengthen technical and political dialogues" related to governance and capacities of AI in the region (Declaración de Montevideo, 2024). The implementation of this intergovernmental framework involves 17 representatives from the region's states, with notable absences from Argentina, Costa Rica, and many Caribbean countries, despite their participation in the first summit.

Second, AI governance was also promoted by creating new multilateral frameworks that involve states based on their status and interests in the development of AI. In this context, a noticeable pattern emerged, characterized by the launch of initiatives by a limited number of like-minded countries and the subsequent inclusion of other relevant partners on an ad hoc basis. Participation was more flexible and generally not limited to specific geographical areas, but depended on the status and interests of countries concerning AI issues.

For example, the final declarations of AI Summits were subscribed to only by the representatives of state governments. The Bletchley Declaration was supported by 28 participants, including the U.S., U.K., China, India, and the European Union (AI Safety Summit, 2023). Another AI Action Summit (2025) statement adopted in Paris in February 2025 was endorsed by more than 60 countries, including China, India, Brazil, the AU Commission, and the EU, but refused by the US and UK due to mentions of "inclusive and sustainable" AI (Bristow, 2025). Such a slow-expanding approach is visible in the US initiative launched in 2023, as a Political Declaration on the Responsible Military Use of Artificial Intelligence and Autonomy. Its endorsement grew from the initial 47 in 2023 to 58 states by the end of 2024. The US-led initiative involved only states, and following the inaugural meeting in March 2024 with the additional participation of some state observers, established three working groups to develop the declaration in practice (Freedberg, 2024).

Another ad hoc multilateral initiative, launched in October 2023 in parallel with the Bletchley summit, was the Chinese Global AI Governance Initiative announced during the third Belt and Road Forum for International Cooperation. It underlined the principle of a people-centred approach, emphasising "respect other countries' national sovereignty and strictly abide by their laws when providing them with AI products and services" (United Nations General Assembly, 2023a). It also stresses that countries should be the only actors involved in international cooperation on AI, based on the principles of equality for all countries in AI development and governance. This initiative was interpreted as a response to Western-led initiatives aimed at countering China's position in the AI field and global discussions, and as an attempt to establish a framework for countries from the Global South, such as the BRICS and G-77 (Roberts, 2024). In practice, the participation in this initiative via dedicated workshops involved around 40 states, mainly from the Global South, but the complete list was not disclosed officially. At the same time, the accompanying Group of Friends on International Cooperation on AI Capacity-building attracted the participation of representatives from around 100 states.

## 5. States' roles in restricted multilateral frameworks

Most AI multilateral frameworks involve different stakeholders, yet they often only anticipate closed intergovernmental processes. Multistakeholder consultations are the typical initial and informal stage in the elaboration of collective state-led AI commitments. They might lead to the drafting of summaries of consultations and preliminary drafts. However, states lead the process from agenda-setting to negotiation and ultimately, the implementation of commitments. The example below illustrates that at each step, the states involved in the multilateral frameworks had the decisive voice.

Firstly, the agenda for debates, including the establishment of multistakeholder processes, was promoted by governments. The progress of debates on AI governance within the G7 framework was driven by the leading role of countries such as Japan, Italy, and Canada, which included this topic in their agendas and advanced discussions during their G7 presidencies. The G7 began its work on AI as early as 2016, with an ICT ministerial debate under the Japanese presidency, and continued in 2017 under the Italian presidency, accompanied by multistakeholder events. This approach was consolidated in 2018 under Canada's presidency with the G7 Innovation Ministers' (2018) Statement on Artificial Intelligence and the G-7 Leaders’ (2018) Charlevoix Common Vision for the Future of AI, followed by a multistakeholder conference on AI with experts representing mostly G-7 members.

In this context, Canada and France played a leading role in the newly established framework, the Global Partnership on AI, which was launched in 2020 by 14 states in North America, Europe, and Asia, along with the European Union. However, it also shows that the state's role can be adverse. Despite that GPAI expanded to 29 members, its operations were strained by the dominance of France and Canada in the organizational architecture, budgetary uncertainty, and an unclear mandate (Wycankoff, 2024). As a result, the GPAI decided in 2024 to reinforce its already existing institutional integration with the OECD’s work on AI (GPAI Secretariat, 2024).

Another example of deliberate state-led agenda-setting involving other stakeholders also occurred in the OECD. The OECD Recommendation was prepared and negotiated under South Korea's chairmanship for both the OECD Expert Group on AI and the OECD Committee on Digital Economy Policy. The informal group of experts, nominated by governments, as well as to some extent industry, civil society, trade unions, the technical community, and academia, drafted an input document that was further elaborated and adopted by intergovernmental bodies. The chairman guided the debates to develop a framework for regulating emerging technology that would be acceptable to the United States[1]. Similarly, governmental leadership is evident in the AI Summits organized in Bletchley (2023), Seoul (2024), and Paris (2025). The British government launched the first summit to consider the consequences of the most advanced AI foundational models of security. All these events, which qualified as multistakeholder, given the presence of industry, international organizations, academia, and civil society, drew a clear line between the leading roles of the British, Korean, and French governments in their preparations and the invited ad hoc roles of other actors.

[1] An anonymous interview with OECD governmental representative, 13 June 2025.

Second, the states are responsible for negotiating a final commitment that summarizes the outcome of any multilateral processes, even if it consistently involves many other stakeholders. For example, REAIM involved around 80 countries that sent delegations to these meetings, as well as multi-stakeholder debates engaging representatives from international organizations, industry, civil society, and academia. During the first summit, more than 2,000 delegates participated in various events, including experts' debates and discussions accompanying the intergovernmental segment. However, the Netherlands was responsible for preparing and negotiating the collective commitment already before the summit in The Hague (Interview, 2023)[2]. Similarly, the preparation through regional consultations among state representatives and the organization of the second summit in South Korea in 2024 reflected a similar pattern. The adopted document, Blueprint for Action (REAIM, 2024), was also drafted by the Korean host following intergovernmental consultations. Another example is that the Council of Europe's preparatory work on the Convention involved many different stakeholders, including experts, civil society organizations, and the private sector. However, the most sensitive stage of the negotiation was conducted behind closed doors among states at the request of the United States, which was unwilling to disclose the states' positions publicly as observers were excluded from this key stage (Bertuzzi, 2023).

Thirdly, most collective AI commitments indicate that the responsibilities for their implementation are intergovernmental and clearly distinguishable from any commitments adopted by non-state actors. For example, the G-7 Hiroshima Process adopted international non-legally binding guiding principles and a code of conduct for organizations developing advanced AI systems in 2023 (G7 Hiroshima Process, 2023a, 2023b). These commitments aim to promote "safe, secure, and trustworthy AI worldwide" when created by private and public organizations, and to apply these standards to "all AI actors" when implemented by different countries (G-7 Hiroshima Process, 2023a). Yet, the G7 indicated that applying the Hiroshima principles and code of conduct can depend on the fact that "different jurisdictions may take their unique approaches to implementing these actions in different ways" (G7 Hiroshima Process, 2023a, 2023b). A similar distinction was observed during the AI Seoul Summit, where the so-called Seoul AI Business Pledge was adopted as a voluntary commitment by a select group of leading Korean and global tech AI companies, alongside two separate declarations from state leaders and ministers (AI Seoul Summit, 2024).

[2] An anonymous interview with a diplomat, 19 December 2023.

Similarly, the ASEAN (2024) issued guidelines on AI Governance and Ethics, which were expanded to include the field of Generative AI a year later (ASEAN, 2025). These guidelines also emphasized that they do not supersede any member states' laws and regulations. The 2024 Guide encourages alignment within ASEAN and fosters the interoperability of AI frameworks "across jurisdictions" (ASEAN, 2024: 1), as it is emphasized that "developers and deployers need to adhere to applicable national laws and regulations" (ASEAN, 2024: 9). The document provides national-level policy recommendations for promoting AI sector development and regional-level policy recommendations for institutionalizing ASEAN intergovernmental collaboration through a dedicated ASEAN AI Working Group on AI Governance.

## 6. Universal AI multilateralism in the UN system

The restricted AI multilateralism reflected varying levels of state involvement depending on their economic status and interest in AI governance. Universal state membership in the UN's core bodies does not appear to limit states' participation in AI governance initiatives. However, the practice reveals varying degrees of state involvement in AI-related initiatives, not only due to their differing interests in AI, but also due to the UN procedural rules and different diplomatic capacities. In this context, it is notable which countries opposed attempts to universalize AI multilateralism within the UN framework, as well as the varying allegiances of different countries.

On the one hand, as shown in Table 2, there is consistent opposition from Russia, Iran, the DPRK, and Belarus, among others, to the adoption of various AI-related UN GA resolutions. This behaviour was also visible during the negotiations of the UNECSO Recommendations when Iran and Russia consistently attempted to obstruct the intergovernmental negotiations (Natorski, 2024). On the other hand, the two UN GA resolutions promoted by the United States (UN GA, 2024c) and China (UN GA, 2024b) were initially sponsored by European and other developed countries, as well as a few African and Asian countries, on the one hand, and by developing countries from Africa, Asia, and the Middle East, on the other hand. Even if finally adopted without a vote and sponsored by many other countries worldwide, they led to the establishment of two Groups of Friends, led by the USA and Morocco, and China and Zambia, with around 70 and 80 participating countries, respectively. This split reveals a divergence among countries along traditional regional UN groups, with a tendency to split between Global North and Global South countries.

In the UN-based universal AI multilateralism, states contributed to the agenda-setting process in UN bodies. Still, a very active role was played by UN system bureaucracies, particularly the UN Secretary-General and many agencies working on practical development, deployment, and standardization of AI in their everyday activities (United Nations System. Chief Executives Board for Coordination, 2024). In this context, Artificial Intelligence emerged as a distinctive topic during the annual UN General Assembly's general debates, amid a growing focus on digital technologies. For instance, digital technologies were mentioned 47 times in 2017 and 94 times in 2023 during the UN GA General Debates. In the 2023 UN GA General Debate, AI was already mentioned in approximately 39 speeches (a rise from only two mentions in 2017), which emphasized the importance of developing human-centred AI based on ethical principles (DigWatch, 2023). Subsequently, the different UN GA resolutions and UN SC activities were initiated only by countries (Table 2). In some cases, they link their leadership with restricted multilateral frameworks. For example, the UN GA resolution on AI in military domains was launched by the Netherlands and South Korea, given their leading role in the organization’s REAIM summits. The involvement of other stakeholders is usually undisclosed. However, in two cases, the UN bodies involved other stakeholders in a parallel process of informal consultations.

Firstly, the elaboration of the Pact of the Future involved extensive informal written and participatory consultations with various stakeholders, as well as UN Member States (for example, the states co-facilitated eight thematic Deep Dives of GDC, including AI, with the industry, non-state actors, and experts) (Teleanu, 2025). Secondly, the elaboration of the UNESCO Recommendations also involved different stakeholders, including an advisory group of experts who prepared the preliminary draft. Yet, in both cases, after multi-stakeholder consultations, the elaboration of commitments was based on exclusive intergovernmental consultations (UNESCO, 2019; UN GA, 2023b Decision). As a result, the first universal instrument addressing the regulation of AI, the UNESCO Recommendation on the Ethics of AI, was adopted after intergovernmental multilateral negotiations by all UNESCO member states (at that time, the US was an observer) in 2021, facing strong opposition from Russia, Iran, and other countries that defended a sovereignist approach to regulating AI against a multilateral approach promoted by predominantly Western countries (Natorski, 2024). Similarly, the negotiation of the Pact of the Future, including the Global Digital Compact, involved several iterations of intergovernmental negotiations of amendments until the very day of its approval. Both the intergovernmental process and stakeholder consultations were maintained as separate streams of work. The intergovernmental process focused on the formal consultations and negotiations of the draft. In contrast, multi-stakeholder consultations were limited to updates and exchanges of views on the drafts (Teleanu, 2025). The bone of contention was preserving the key role for member states in relations to other stakeholders in AI governance. The G77 and China (2023: 1-2) emphasized that the GDC "must be a States-led process and ensure full and equal participation of all States, especially developing countries" while "there is a strong need that stakeholders observe national laws, regulations, principles, rules and norms for their responsible behavior in ICT environment".

Finally, as in the case of restricted multilateralism, the commitments adopted by the UN General Assembly show the leading responsibility of states for implementation in a multi-stakeholder context. Therefore, in the GDC, it is the governments' responsibility to implement it at national, regional, and global levels, while "taking into account different national realities, capacities, and levels of development, and respecting national policies and priorities and applicable legal frameworks" (United Nations General Assembly, 2024a: Art. 64). At the same time, the government recognize the fundamental role of different stakeholders and pledged to "strengthen our collaboration and leverage multi-stakeholder cooperation to achieve the objectives set out in this Compact" (United Nations General Assembly, 2024a: Art 65). Yet, this commitment is accompanied by the emphasis on "their respective roles and responsibilities". Despite these caveats, the multistakeholder aspect is particularly prominent in the context of AI governance, with "full and equal representation of all countries, especially developing countries, and the meaningful participation of all stakeholders" (United Nations General Assembly, 2024a, Art. 50).

A similar distinction between governments and other actors appears in other UN GA resolutions. For example, the UN GA resolution on the capacity-building of AI, promoted by China, calls on states to include this topic in their national policies and strengthen cooperation. Meanwhile, other stakeholders could support this state-led effort mainly through non-state collaboration and providing technical and financial resources (United Nations General Assembly, 2024b). The UN GA resolution on safe, secure, and trustworthy AI, promoted by the US, takes a different approach by addressing states and "inviting" other stakeholders to contribute to regulatory, governance, promotional, and capacity-building efforts within their respective roles and responsibilities. While the resolution emphasizes intergovernmental cooperation, it also adopts a multi-stakeholder approach and urges the private sector to comply with applicable international and domestic laws (United Nations General Assembly, 2024c). However, as noted, the impact of this resolution might be limited by the precedence of national policies, and the possibility of states opting out "under the pretext of their own national strategies and requirements" (Knauer, OXIO 792). Additionally, as noted by several civil society organizations, "the framing of AI governance proposed does not reflect a true multistakeholder model and could be stronger when it comes to meaningful participation and inclusivity" (Accessnow, 2024).

## 7. Conclusion

It has become a convention to discuss AI "governance" and a "multi-stakeholder" approach, encompassing the involvement of various actors in AI-related matters. International Organizations, private technological companies, non-governmental organizations (including associations, networks, and civil society groups), academic institutions, and individual experts actively try to shape the discourse on AI. As a result, it might seem that the hierarchies among private and public entities are fluid and appear to challenge the traditionally dominant role of states in multilateralism. In this context, this chapter aimed to answer the question of how multilateralism addressed the emergence of the general-purpose technology of AI and how it affected the roles of states.

The analysis revealed that states continue to define the current landscape of AI global governance. States retain ultimate authority over other actors and commitments that emerge from these frameworks. They promote and lead new initiatives, shape them, even veto them, control the implementation of commitments, and finally decide on the inclusion of other non-state actors. Non-state actors are typically involved in the consultation stage. At the same time, the involved non-state actor could not neglect these gate-keeping roles of governments and act beyond the margins of the analysed AI governance framework.

The analysis revealed these states' roles following a strikingly convergent approach to defining the guiding principles of multilateralism in the context of AI. The principles are constituted around the notions of epochalism, determinism, and dialectics. They respectively define AI as a general-purpose technology that disrupts historical trajectories, embodies its inevitable character, and fosters a dialectical understanding that embraces the contradiction of risks and opportunities. In this context, the chapter illustrates the two ways in which multilateralism adapts to the emergence of AI.

First, many existing multilateral frameworks, both thematic and regional, have incorporated AI topics into their agendas. G-7 and OECD played a pioneering role in establishing key commitments for the developed economies, but several other regional bodies also followed (mainly the EU and the Council of Europe). Notably, the core initiatives of AI-transformed multilateralism develop within existing restricted frameworks yet remain open to including interested non-member states. Nonetheless, besides established frameworks, the states also launched several ad-hoc initiatives (mainly the AI Action Summit, GPAI, and REAIM), embracing mostly like-minded countries. Under the leadership of middle power states (Canada, the UK, South Korea, the Netherlands, and France), the membership of states in these initiatives evolves and tends to be open to non-Western states, including China and other emerging powers.

Second, the existing multilateralism does not equally involve all states and regions worldwide. The initiatives at the regional level in Asia, Africa, and Latin America are institutionalized following Western and UN-led initiatives. Whether work of the ASEAN, African Union, or ad-hoc grouping of Latin American and Caribbean states focusing on the ethics of AI will consolidate enough remains to be seen. Yet, it is already significant that Global South countries are increasingly pushing to establish their position in the emerging global governance landscape for AI by establishing regional multilateral frameworks. However, it also partially explains the relatively few concrete results and commitments of the universal multilateralism around the United Nations system. The scope of the agreements reached indicates that the UN system serves only as an additional layer for AI multilateralism, co-existing with restricted multilateralism, given the different relevance attributed to them by states. The UN framework visualized that the opportunism of the US and China could lead to some tangible results. Still, even though both AI world leaders selectively and reluctantly might promote AI multilateralism, they must face the sovereignist claims of Russia, Iran, and a few others, which reject multilateral collaboration on the governance of AI technology, and the indifference of many other UN member states.

Although research on the global governance of AI is still in its early stages, it must address many gaps to gain a deeper understanding of its evolving structure. While this chapter has argued that the role of states remains vital in AI multilateralism, it is also essential to better understand the noticeable role of non-state actors in these multistakeholder processes. Although they are not equal to states in terms of AI governance, their consistent presence in these frameworks requires studying their motivations and interests. The authority of states as the ultimate arbiters of the global governance framework might be indirectly influenced by industry, international organizations, civil society, experts, and academia. It can also include studies on how participation in AI technology affects a state's status and authority in international relations. Still, further study is required to fully understand the state's role throughout the entire policy cycle. Future research can focus primarily on how stakeholder involvement helps to learn about the technology itself, influences state preferences, and legitimizes state commitments. Another fruitful line of research can be to analyze the role of bureaucracies in international organizations, particularly the prolific UN system, as agenda-setters and brokers of international agreements, as well as the role of advisory expert bodies that establish the grounds for future interstate interactions. Similarly, it remains a question whether private non-public actors, in particular companies, can create, on the margin of the state hierarchical authority, a sustainable private multilateral framework of cooperation in the context of relentless capitalist competition for AI market domination.

Another future line of research emerging from the current state of AI multilateralism is the emergence and consolidation of specific AI-related norms and rules that define the standard of appropriateness for this technology. Although non-legally binding commitments dominate the current landscape, they establish standards that the involved partners assume as their national commitment. This chapter illustrates that it remains the state's competence to translate international soft-law commitments into legal frameworks within its jurisdiction. Therefore, it remains crucial to better understand the multi-level connections between global and regional, on the one hand, and state and non-state, on the other hand, regulations governing AI technology.

At this moment, it can be observed that AI multilateralism is agreed upon in the shadow of the state hierarchy as the ultimate authority. However, the current state of AI global governance presents numerous opportunities for multidisciplinary research that can benefit from robust empirical data collection and analysis. It also highlights the chance to revisit many traditional IR concepts and theories in light of this general-purpose technology, as well as to develop new insights and ideas that reflect the technology-driven changes in international politics.

**Acknowledgements**

The research for this chapter greatly benefited from the two research grants: NWO (Dutch Research Council) Science Diplomacy Fund – Embassy Science Fellowship (2023-2025) for the project "Artificial Intelligence: geopolitical and security-related implications of new technology in South Korea" and Maastricht University SBE Incentive Grant (2024-2030) for the project "Global Regime Complex System of AI Policies" which allowed to conduct field work in the Republic of Korea and collect relevant data from other site of multilateral meetings on AI.

**Table 2. UN key activities on AI in intergovernmental organs**

| Activity | Date | Promotor | Voting | Role of non-state stakeholders |
|---|---|---|---|---|
| **UN General Assembly** | | | | |
| A/RES/78/241- Lethal autonomous weapons systems | 22 December 2023 | Austria and 25 co-sponsors | Yes – 152<br>No – 4 (Belarus, India, Mali, Russia)<br>Abstain – 11 (China, DPRK, Iran, Isreal, Madagascar, Niger, Saudi Arabia, South Sudan, Syria, Türkiye, UEA). | Support of non-state stakeholders during external debates |
| A/RES/78/265 - Seizing the opportunities of safe, secure and trustworthy artificial intelligence systems for sustainable development | 21 March 2024 | United States | Adopted without vote | Unknown |
| A/RES/78/311 - Enhancing international cooperation on capacity-building of artificial intelligence | 1 July 2024 | China | Adopted without vote | Unknown |
| A/RES/79/1 - The Pact for the Future (and Annex I Global Digital Compact) | 22 September 2024 | UN Secretary General and Member States | Adopted without vote, with reluctance of Russia, Belarus, DPRK, Iran, Nicaragua, and Syria (last minute attempt to stop the adoption) | Extensive open consultations states and stakeholders in 2023 and 2024 |
| A/RES/79/239 Artificial intelligence in the military domain and its implications for international peace and security | 24 December 2024 | The Netherlands and the Republic of Korea | Yes - 159<br>No – 2 (Russia and DPRK)<br>Abstain – 5 (Belarus, Ethiopia, Iran (Islamic Republic of), Nicaragua, Saudi Arabia) | Unknown |
| A/RES/79/62 - Lethal autonomous weapons systems | 2 December 2024 | Austria and 15 countries | Yes – 166<br>No – 3 (Belarus, DPRK, and Russia).<br>Abstain – 15 (China, Estonia, Fiji, India, Iran, Israel, Latvia, Lithuania, Nicaragua, Poland, Romania, Saudi Arabia, Syria, Türkiye, and Ukraine.) | Support of non-state stakeholders during external debates |
| **UN Security Council** | | | | |
| S/PV.9381 - Briefing on Artificial intelligence: opportunities and risks for international peace and security | 18 July 2023 | United Kingdom | No voting | Two experts |
| Arria-formula Meeting on Artificial Intelligence: Its Impact on Hate Speech, Disinformation and Misinformation | 19 December 2023 | Albania and the United Arab Emirates | No voting | Two experts |
| S/PV.9821 Briefing on Maintenance of international peace and security: Artificial intelligence | 19 December 2024 | United States | No voting | Two experts |
| Harnessing safe, inclusive, trustworthy AI for the maintenance of international peace and security | 4 April 2025 | Greece, France, Republic of Korea, Armenia, Italy, the Netherlands | No voting | Three experts |

Source: Own elaboration